\documentclass[pra,aps,twocolumn,epsfig,superscriptaddress,showpacs]{revtex4}
\usepackage{amsmath}
\usepackage{amssymb}
\usepackage{graphicx}
\usepackage{dcolumn}
\usepackage{bm}
\usepackage[colorlinks=false,dvipdfm]{hyperref} 
\usepackage{setspace}
\newcommand{\tr}{\textrm{Tr}}

\newcommand{\bra}[1]{\ensuremath{\langle#1|}}

\newcommand{\ket}[1]{\ensuremath{|#1\rangle}}

\newcommand{\BE}{\begin{equation}}
\newcommand{\EE}{\end{equation}}
\newcommand{\kommentar}[1]{}

\newcommand{\be}{\begin{equation}}
\newcommand{\ee}{\end{equation}}

\begin{document}

\author{Zhihao Ma}
\affiliation{Department of Mathematics, Shanghai Jiaotong
University, Shanghai, 200240, P.R.China}

\author{Fu-Lin Zhang}
\affiliation{Theoretical Physics Division, Chern Institute of
Mathematics, Nankai University, Tianjin, 300071, P.R.China}

\author{Zhi-Hua Chen}
\affiliation{Department of Science, Zhijiang college, Zhejiang
University of technology, Hangzhou, 310024, P.R.China}

\author{Jing-Ling Chen}
\email[Email:]{chenjl@nankai.edu.cn}\affiliation{Theoretical Physics
Division, Chern Institute of Mathematics, Nankai University,
Tianjin, 300071, P.R.China}

\date{\today}

\begin{abstract}

We report a new metric of quantum states. This metric is build up
from super-fidelity, which has deep connection with the
Uhlmann-Jozsa fidelity and plays an important role in quantifying
entanglement. We find that the new metric possess some interesting
properties.

\end{abstract}

\title{ Super fidelity and related metrics }
\pacs{03.67.-a, 03.65.Ta} \keywords{Super-Fidelity; Metric; Quantum
state} \maketitle

\section{Introduction}

In quantum information theory, a fundamental task is to distinguish
two quantum states. One of the main tools used in distinguishability
theory is trace metric, another closed related tool is quantum
fidelity \cite{Niebook,Fid1}. Both are widely used by the quantum
information science community and have been found applications in a
number of problems such as quantifying entanglement
\cite{97Vedral2275,98Vedral1619}, quantum error correction
\cite{08Kosut020502}, quantum chaos \cite{Zana1}, and quantum phase
transitions \cite{Wang1}.

Suppose  $\rho$ and $\sigma$ are two quantum states, then the
Uhlmann-Jozsa fidelity \cite{Niebook,Fid1} between $\rho$ and
$\sigma$ is given by
\begin{eqnarray}F(\rho, \sigma)=[\tr
\sqrt{\rho^{\frac{1}{2}}\sigma\rho^{\frac{1}{2}}}]^{2}
\end{eqnarray}

We know that for the case of qubits, Uhlmann-Jozsa fidelity has a
simple form. From the Bloch sphere representation of quantum states,
a qubit is described by a density matrix as:
\begin{eqnarray}\rho(\textbf{u})=\frac{1}{2}(\textbf{I}+\sigma\cdot
\textbf{u})
\end{eqnarray}
where $\textbf{I}$ is the $2\times 2$ unit matrix and
$\sigma=(\sigma_{1},\sigma_{2},\sigma_{3})$ are the Pauli matrices.
Assume $\rho(\textbf{u})$ and $\rho(\textbf{v})$ are two states of
one qubit, then they can represented by two vectors $\textbf{u}$ and
$\textbf{v}$ in the Bloch sphere. The Uhlmann-Jozsa fidelity for
qubits has an elegant form:
\begin{eqnarray}
F(\rho(\textbf{u}), \rho(\textbf{v}))=\frac{1}{2}[1+\textbf{u}\cdot
\textbf{v}+\sqrt{1-|\textbf{u}|^{2}}\sqrt{1-|\textbf{v}|^{2}}]
\end{eqnarray}
where $\textbf{u}\cdot \textbf{v}$ is the inner product of
$\textbf{u}$ and $\textbf{v}$, and $|\textbf{u}|$ is the magnitude
of $\textbf{u}$.

We know that for general quantum states, the Uhlmann-Jozsa fidelity
has no simple form like the case of qubits. To use the simple form
of fidelity, we note that in \cite{Mis}, the authors introduce a new
fidelity, called super-fidelity, defined as
\begin{eqnarray}G(\rho_1, \rho_2):=\tr \rho_1 \rho_2 + \sqrt{(1-\tr \rho_1^2)(1-\tr
\rho_2^2)}\end{eqnarray} and it was proved that when $\rho_1$ and
$\rho_2$ are two qubits, super-fidelity $G(\rho_1, \rho_2)$
coincides with Uhlmann-Jozsa fidelity $F(\rho_1, \rho_2)$.

The super-fidelity $G(\rho_1, \rho_2)$ has some appealing
properties\cite{Mis,Puch,Mend}. Let
$\rho_{u}=\frac{1}{N}(I+\sqrt{\frac{N(N-1)}{2}}\overrightarrow{\lambda}.\textbf{u})$
be the density matrix of a qunit($N\times N$ quantum state), where
$I$ is the $N\times N$ unit matrix,
$\overrightarrow{\lambda}=(\lambda_{1},\lambda_{2},...\lambda_{N^2-1})$
are the generators of $SU(N)$, and $\textbf{u}$ is the
$(N^2-1)$-dimensional Bloch vector. Then super-fidelity can be
rewritten as $G(\rho_{u}, \rho_{v})=\frac{1}{N}[1+(N-1)\times
\textbf{u}.\textbf{v}+(N-1)\times\sqrt{(1-|\textbf{u}|^{2})(1-|\textbf{v}|^{2})}]$.
This shows that super-fidelity only depends on the magnitudes of
$\textbf{u}, \textbf{v}$ and the angle between them(that is,
$\textbf{u}.\textbf{v}$). This property make super-fidelity easy to
calculate, and has a clear geometrical interpretation.

Moreover, very recently, it was found that super-fidelity play an
important role in quantifying entanglement \cite{Ma1616}. So it is
natural to study the property of super-fidelity in further step.

Recall that super-fidelity by itself is not a metric. It is a
measure of the ``closeness'' of two states. If we say a function
$d(x,y)$ defined on the set of quantum states is a metric, it should
satisfies the following four axioms:

(M1). $d(x,y) \ge 0$ for all states $x$ and $y $;

(M2). $d(x,y)=0$ if and only if  $x=y$;

(M3). $d(x,y)=d(y,x)$ for all states $x$ and $y$;

(M4). The triangle inequality: $d(x,y)\leq d(x,z)+d(y,z)$ for all
states $x, y$ and $z$.

For super-fidelity, one can define the following three functions\cite{Mend}:\begin{eqnarray}A(\rho,\sigma)&:=&\arccos{\sqrt{G(\rho,\sigma)}},\\
B(\rho,\sigma)&:=&\sqrt{2-2\sqrt{G(\rho,\sigma)}},\\
C(\rho,\sigma)&:=&\sqrt{1-G(\rho,\sigma)},
\end{eqnarray}

It was proved in \cite{Mend} that $C(\rho,\sigma)$ is a genuine
metric, that is, it satisfying the axioms M1-M4, while
$A(\rho,\sigma)$ and $B(\rho,\sigma)$ do not preserve the metric
properties.

The purpose of this paper is to introduce a novel method to define
metric of quantum states based on super-fidelity. Surprising, we
find the metric induced by the new method coincides with the metric
introduced in \cite{Mend} for the qubits case, and the new metrics
have deep connection with spectral metric. Also we find the new
metrics possess some appealing properties which make the metrics
very useful in quantum information theory. The paper is organized as
follows: In Sec. II, two new metrics were defined, and the metric
character of the metrics were established. In Sec. III, intrinsic
properties of the two metrics were discussed. Conclusion and
discussion were made in the last section.

\section{Metric induced by super-fidelity}

The most widely used metric may be trace metric, which was defined
as
\begin{eqnarray}D_{tr}(\rho, \sigma)=\frac{1}{2}\tr|\rho-
\sigma|\end{eqnarray} On the other hand, one can define other types
of distance measures for quantum states, and these measures also
have their own advantages, see \cite{Niebook, Mis, Mend, 95Fuchs,
Nie062310,Ra06, Ra02,Ma064325,Ma1616,Ma3407}.

Let us define a new metric of states as follows:
\begin{eqnarray}D_{G}(\rho,
\sigma)=\max\limits_{\tau}|G(\rho,\tau)-G(\sigma,\tau)|\label{G-metric}
\end{eqnarray}
where the maximization is taken over all quantum states $\tau$ (mix
or pure). We call this metric $D_{G}(\rho, \sigma)$ as the G-metric,
and the state $\tau$ that attained the maximal is called the optimal
state for the metric $D_{G}(\rho, \sigma)$.

The above definition of metric may be not easy to calculate. So we
can change its definition slightly. If $\tau$ is a pure state, then
super-fidelity can be simplified as $G(\rho,\tau)=\tr(\rho \tau)$,
hence one can define another version of metric as follows:
\begin{eqnarray}D_{PG}(\rho,
\sigma)=\max\limits_{\tau}|G(\rho,\tau)-G(\sigma,\tau)|\label{PG-metric}
\end{eqnarray}
where the maximization is taken over all pure states $\tau$. We call
this metric $D_{PG}(\rho, \sigma)$ as the PG-metric, and call the
pure state $\tau$ that attained the maximal as the optimal pure
state.

First we consider the case of qubits.

\emph{Proposition 1\cite{Ma3407}.} For the qubit case,
$D_{PG}(\rho,\sigma)$ equals to the trace metric, namely
$D_{PG}(\rho,\sigma)
=D_{tr}(\rho,\sigma)=\frac{1}{2}\tr|\rho-\sigma|$.

We can connect our metric with the metric introduced in \cite{Mend}
as following:

\emph{Proposition 2\cite{Ma3407}.} For the qubit case, $D_{G}(\rho,
\sigma)=C(\rho,\sigma)=\sqrt{1-G(\rho, \sigma)}$.

Now we come to discuss the case of qunit(i.e., $N\times N$ quantum
states). In this case, if $\tau$ is a pure state, then the
super-fidelity has a simple form: $G(\rho,\tau)=\tr(\rho \tau)$,
this make the PG-metric easy to study. So we first show the metric
character of $D_{PG}(\rho, \sigma)$, where the optimal state $\tau$
is restricted to pure state, and then turn to show the metric
character of $D_{G}(\rho, \sigma)$.

We need the following concepts: For two quantum state $\rho$ and $
\sigma$, let $\lambda_{i}$, $(i=1, 2, 3, ... , n)$, be all
eigenvalues of $\rho-\sigma$, and arranged as $\lambda_{1}\geq
\lambda_{2}\geq...\geq \lambda_{n}$. Define $E(\rho, \sigma):=\max
\lambda_{i}$. We can give an interpretation of $E(\rho, \sigma)$ as
follows: Let $\rho$ and $\sigma$ be two quantum states, then the
following is well known(for example, see \cite{Horn91}):
\begin{eqnarray}
E(\rho, \sigma)=\max\limits_{\tau}\tr[\tau(\rho-\sigma)],
\end{eqnarray}
where the maximization is taken  over all pure states $\tau$.

Note that generally $E(\rho, \sigma)$ is not a metric, since
$E(\rho, \sigma)$ may not equal to $E(\sigma, \rho)$, but we can
symmetrize it as:
\begin{eqnarray}
D_{S}(\rho, \sigma):=\max[E(\rho, \sigma),E(\sigma, \rho)]=\max
|\lambda_{i}|
\end{eqnarray}
where $|\lambda_{i}|$ is the absolute value of $\lambda_{i}$. From
the knowledge of matrix analysis, we get that $D_{S}(\rho, \sigma)$
equal to the spectral metric between $\rho$ and $\sigma$, which was
defined as the largest singular value of $\rho-\sigma$, hence we
know that $D_{S}(\rho, \sigma)$ is in fact the spectral metric.
Moreover, we have the following:

\emph{Proposition 3 \cite{Ma3407}.} For quantum states $\rho$ and
$\sigma$, $D_{PG}(\rho,\sigma)=D_{S}(\rho,\sigma)$, that is, the
PG-metric is nothing but the spectral metric.

Now we know that the PG-metric is in fact the spectral metric, so it
is a true metric. In the following we shall prove that the G-metric
is also a true metric.

\emph{Theorem 1.} The T-metric $D_{G}(\rho, \sigma)$ as shown in Eq.
(\ref{G-metric}) is truly a metric, i.e., it satisfies conditions
M1-M4.

\emph{Proof.} From the definition, it is easy to prove conditions M1
and M3 hold. What we need to do is to prove conditions M2 and M4. If
$\rho=\sigma$, then of course $D_{G}(\rho, \sigma)=0$. If
$D_{G}(\rho, \sigma)=0$, we will prove $\rho=\sigma$. From the
definition, we know that $D_{G}(\rho, \sigma)\geq D_{PG}(\rho,
\sigma)$, so we get $D_{PG}(\rho, \sigma)=0$, since $D_{PG}(\rho,
\sigma)$ is a true metric, we get $\rho=\sigma$. Now we come to
prove M4, the triangle inequality $D_{G}(\rho, \sigma)\leq
D_{G}(\rho, \tau)+ D_{G}(\sigma, \tau)$. $D_{G}(\rho,
\sigma)=\max\limits_{\tau}|G(\rho, \tau)-G(\sigma, \tau)|$, and
suppose $\tau$ is the optimal state that attains the maximal, so
$D_{G}(\rho, \sigma)=|G(\rho, \tau)-G(\sigma, \tau)|$. Assume that
$|G(\rho, \tau)-G(\sigma, \tau)|= G(\rho, \tau)-G(\sigma, \tau)$,
then we get $G(\rho, \tau)-G(\sigma, \tau)= G(\rho,
\tau)-G(w,\tau)+G(w,\tau)-G(\sigma, \tau)\leq |G(\rho,
\tau)-G(w,\tau)|+|G(w,\tau)-G(\sigma, \tau)|\leq D_G(\rho,w)+
D_G(w,\sigma)$. Thus one finally has $D_G(\rho, \sigma) \leq
D_G(\rho,w)+ D_G(\sigma, w)$. Theorem is proved.

\section{Properties of $D_{G}$ and $D_{PG}$}

We know that for qubits, $D_{G}$ has a clear form as: $D_{G}(\rho,
\sigma)=\sqrt{1-G(\rho,\sigma)}$, how about higher dimension?

For the qunit case, one does not have the relation $D_G(\rho,
\sigma) =\sqrt{1-G(\rho,\sigma)}$ as in Proposition 2.

However, the following upper bound holds: For qunits $\rho$ and
$\sigma$, the following relation holds:
\begin{eqnarray}D_{G}(\rho,\sigma) \leq \sqrt{\frac{2\times
(N-1)}{N}}\times\sqrt{1-G(\rho,\sigma)}
\end{eqnarray}

Proof: Let $\rho=\rho(\textbf{u})$, $\sigma=\sigma(\textbf{v})$ and
$\tau=\tau(\textbf{w})$, where $\textbf{u},\textbf{v},\textbf{w}$
are the corresponding Bloch vectors of the states $\rho,
\sigma,\tau$, then one obtains
\begin{eqnarray}
&&|G(\rho,\tau)-G(\sigma,\tau)|\times\frac{N}{N-1} \nonumber\\
&&=\biggr|(\textbf{u}-\textbf{v})\cdot
\textbf{w}+\sqrt{1-|\textbf{w}|^{2}}(\sqrt{1-|\textbf{u}|^{2}}-\sqrt{1-|\textbf{v}|^{2}})\biggr|\nonumber\\
&&\leq|\textbf{u}-\textbf{v}||\textbf{w}|+
\sqrt{1-|\textbf{w}|^{2}}\;|\sqrt{1-|\textbf{u}|^{2}}-\sqrt{1-|\textbf{v}|^{2}}|\nonumber\\
&&\leq\sqrt{|\textbf{u}-\textbf{v}|^2+
|\sqrt{1-|\textbf{u}|^{2}}-\sqrt{1-|\textbf{v}|^{2}}|^2}\\
&&=\sqrt{[2-2\textbf{u}.\textbf{v}-2\sqrt{1-|\textbf{u}|^{2}}\sqrt{1-|\textbf{v}|^{2}}]}\nonumber\\
&&=\sqrt{\frac{2\times
N}{N-1}}\times\sqrt{1-G(\rho(\textbf{u}),\sigma(\textbf{v}))}\nonumber.
\end{eqnarray}

Now we will discuss the inequality (13) in more detail. When $N=2$,
i.e., in the case of qubits, we get that inequality (13) becomes
equality, that is, $D_{G}(\rho, \sigma)=\sqrt{1-G(\rho,\sigma)}$.
But for higher dimension, the equality sign does not hold in
general. Why?

The reason is subtle. When the equality sign holds, i.e.,
$D_{G}(\rho(\textbf{u}),\sigma(\textbf{v}))=\sqrt{\frac{2\times
(N-1)}{N}}\sqrt{1-G(\rho(\textbf{u}),\sigma(\textbf{v}))}$, then the
inequality (14) need to be quality, that means the optimal state
$\tau=\tau(\textbf{w})$ is always attained, where $\textbf{w}$ is a
vector that parallels to $\textbf{u}-\textbf{v}$, and
$|\textbf{w}|=\frac{\sqrt{N-1}|\textbf{u}-\textbf{v}|}{\sqrt{2\times
N}\sqrt{1-G(\rho(\textbf{u}),\rho(\textbf{v}))}}$, but in fact we
can not always get such optimal state. Because such an operator
$\tau(\textbf{w})$ may not be a density operator! We will explain it
in the following.

It is well known that every $N\times N$ density matrix can be
represented by the $(N^2-1)$-dimensional Bloch vector as:
$\rho(\textbf{u})=\frac{1}{N}(I+\sqrt{\frac{N(N-1)}{2}}\overrightarrow{\lambda}.\textbf{u})$,
but the converse is not true, i.e., not all operator of the form
$\frac{1}{N}(I+\sqrt{\frac{N(N-1)}{2}}\overrightarrow{\lambda}.\textbf{u})$
is a density matrix, where $\textbf{u}$ is an arbitrary
$(N^2-1)$-dimensional Bloch vector. Note that a density matrix must
satisfy three conditions: (a). Trace unity,
$\tr(\rho(\textbf{u}))=1$. (b). Hermitian,
$\rho(\textbf{u})^{+}=\rho(\textbf{u})$; and (c). positivity, i.e.,
all eigenvalues of $\rho(\textbf{u})$ are non-negative.

Indeed, the operator
$\frac{1}{N}(I+\sqrt{\frac{N(N-1)}{2}}\overrightarrow{\lambda}.u)$
automatically satisfies the conditions (a) and (b). However, not
every vector $\textbf{u}$, $|\textbf{u}|\leq 1$, allows
$\rho(\textbf{u})$ satisfies the positive  condition (c), for
example, see \cite{Chen054304}.

To get that inequality (14) becomes equality, we need that the
optimal state $\tau=\tau(\textbf{w})$ is a density matrix, where
$\textbf{w}$ is a vector that parallels to $\textbf{u}-\textbf{v}$,
and
$|\textbf{w}|=\frac{\sqrt{N-1}|\textbf{u}-\textbf{v}|}{\sqrt{2\times
N}\sqrt{1-G(\rho(\textbf{u}),\rho(\textbf{v}))}}$, but this is not
always true in general. So we can only get the inequality (13).

The following counterexample will show that strict inequality will
occur.

\emph{Example 1.} Let
$\ket{\psi}=\frac{\sqrt{3}}{2}\ket{00}+\frac{1}{2}\ket{11}$,
$\ket{\phi}=\frac{1}{2}\ket{00}+\frac{\sqrt{3}}{2}\ket{11}$. Define
$\rho=\ket{\psi}\bra{\psi},\sigma=\ket{\phi}\bra{\phi}$, then we get
that  $D_{G}(\rho, \sigma)=\frac{1}{2}$, while $\sqrt{\frac{2\times
(3)}{4}}\times\sqrt{1-G(\rho,
\sigma)}=\sqrt{\frac{3}{8}}>\frac{1}{2}$.

Now we will study the intrinsic properties of the G-metric $D_{G}$
and PG-metric $D_{PG}$. We are interested in the following
properties:

{\bf Property 1: contractive  under quantum operation.} suppose $T$
is a quantum operation, i.e., a completely positive trace preserving
(CPT) map, and $\rho, \sigma$ are density operators, we say a metric
$D(\rho, \sigma)$ is contractive under quantum operation, if the
following holds:
\begin{eqnarray}D(T(\rho), T(\sigma))\leq D(\rho, \sigma)\end{eqnarray}

Why we study the property of contractive  under quantum operation?
It has a physical interpretation \cite{Nie062310}: a quantum process
acting on two quantum states can not increase their
distinguishability.

{\bf Property 2: joint convex property.} we say that the metric
$D(\rho, \sigma)$ has the convex property, if  $p_{j}$ are
probabilities, then \begin{eqnarray}D(\sum_{j}p_{j}\rho_{j},
\sum_{j}p_{j}\sigma_{j})\leq \sum_{j}
p_{j}D(\rho_{j},\sigma_{j})\end{eqnarray}

The joint convex property also has a physical interpretation
\cite{Nie062310}: the distinguishability between the states
$\sum_{j}p_{j}\rho_{j}$ and $\sum_{j}p_{j}\sigma_{j}$, where $p_{j}$
is not known, can never be greater than the average
distinguishability when $p_{j}$ is known.

We know that the Uhlmann-Jozsa fidelity $F(\rho,\sigma)$ has the
{\bf CPT expansive property}: If $\rho$ and $\sigma$ are density
matrices, $\Phi$ is a CPT map, then
\begin{eqnarray}F(\Phi(\rho),\Phi(\sigma))\geq F(\rho,\sigma)\end{eqnarray}

We may guess that the super-fidelity $G(\rho,\sigma)$  also has the
CPT expansive property, the following counterexample shows that this
property does not holds.

\emph{ Example 2 \cite{Mend}.} Let $$A=\left (\begin{array}{cccc} 0 & 0& 0& 0\\
1&0 & 0 & 0\\
0 & 0& 0& 0\\
0&0 & 1 & 0\end{array}\right ), \;\;\;\;\;
B=\left (\begin{array}{cccc} 0 & 0& 0& 0\\
0&1 & 0 & 0\\
0 & 0& 0& 0\\
0&0 & 0 & 1\end{array}\right ),$$

Define $\Phi(\gamma)=A\gamma A^{+}+B\gamma B^{+}$, where $\gamma$ is
an arbitrary density operator, then we defined a completely positive
trace preserving map.

Let $\rho$ and $\sigma$ be the density operators defined by $$\rho=\left (\begin{array}{cccc} \frac{1}{2} & 0& 0& 0\\
0&\frac{1}{2} & 0 & 0\\
0 & 0& 0& 0\\
0&0 & 0 & 0\end{array}\right ), \;\;\;\;\;
\sigma=\left (\begin{array}{cccc} 0 & 0& 0& 0\\
0&0 & 0 & 0\\
0 & 0& \frac{1}{2}& 0\\
0&0 & 0 & \frac{1}{2}\end{array}\right ),$$

Then $$\Phi(\rho)=\left (\begin{array}{cccc} 0 & 0& 0& 0\\
0&1 & 0 & 0\\
0 & 0& 0& 0\\
0&0 & 0 & 0\end{array}\right ), \;\;\;\;\;
\Phi(\sigma)=\left (\begin{array}{cccc} 0 & 0& 0& 0\\
0&0 & 0 & 0\\
0 & 0& 0& 0\\
0&0 & 0 & 1\end{array}\right ),$$

One then easily obtains $G(\rho,\sigma)>G(\Phi(\rho),\Phi(\sigma))$,
which shows that the CP expansive property property does not hold
for super-fidelity.

So we get that the metric $C(\rho,\sigma):=\sqrt{1-G(\rho,\sigma)}$
introduced in \cite{Mend} is  not contractive  under quantum
operation.

However, we can prove the following:

\emph{Theorem 2.} The PG-metric $D_{PG}(\rho, \sigma)$  is
contractive under quantum operation, that is, $D_{PG}(\phi(\rho),
\phi(\sigma))\leq D_{PG}(\rho, \sigma)$.

{\bf Proof.} Suppose $\gamma$ is the optimal pure state for quantum
states $\phi(\rho), \phi(\sigma)$, so we get
$D_{PG}(\phi(\rho),\phi(\sigma))=|G(\phi(\rho),\gamma)-G(\phi(\sigma),\gamma)|=
|(\tr \phi(\rho)\gamma)-(\tr \phi(\sigma)\gamma)|$

Let $\phi$ be a quantum operation, and denote
$\gamma^{'}:=\phi^{*}(\gamma)$. Then we have
\begin{eqnarray*}&&D_{PG}(\phi(\rho), \phi(\sigma))\\
&=&|(\tr \phi(\rho)\gamma)-(\tr \phi(\sigma)\gamma)|\\
&=& |(\tr \rho\phi^{*}(\gamma))-(\tr \sigma\phi^{*}(\gamma))|\\
&=& |(\tr \rho\gamma^{'})-(\tr \sigma\gamma^{'})|\\
&\leq & D_{PG}(\rho, \sigma) \end{eqnarray*} Theorem is proved.

Note that the PG-metric is in fact the spectral metric, and it was
proved in \cite{Pe} that spectral metric is contractive under
quantum operation, here we give an elementary proof, our method is
quite different from that of \cite{Pe}.

How about the G-metric? Numerical experiment shows that the G-metric
$D_{G}(\rho, \sigma)$ is  not contractive  under quantum operation.

Now we discuss the joint convex property.

 \emph{Proposition 4.} (joint convexity of the PG-metric): Let
$\{p_{i}\}$ be probability distributions over an index set, let
$\rho_{i}$ and $\sigma_{i}$ be density operators with the indices
from the same index set. Then
\begin{eqnarray}D_{PG}(\sum_{i}p_{i}\rho_{i}, \sum_{i}p_{i}\sigma_{i})\leq
\sum_{i}p_{i}D_{PG}(\rho_{i}, \sigma_{i})\end{eqnarray}

We know that $D_{PG}(\rho, \sigma)=D_{S}(\rho, \sigma)=\max(E(\rho,
\sigma),E(\sigma, \rho))$, so we only need to prove the following
holds:
\[E(\sum_{i}p_{i}\rho_{i}, \sum_{i}p_{i}\sigma_{i})\leq
\sum_{i}p_{i}E(\rho_{i}, \sigma_{i})\] since $E(\rho,
\sigma)=\max\limits_{\gamma}\tr(\gamma(\rho-\sigma)),$ where the
maximization in the right hand is taken  over all pure states
$\gamma$, then there exists a pure state $\gamma$ such that
\[E(\sum_{i}p_{i}\rho_{i}, \sum_{i}p_{i}\sigma_{i})=\sum_{i}p_{i}\tr(\gamma(\rho_{i}-\sigma_{i}))\leq \sum_{i}p_{i}E(\rho_{i}, \sigma_{i}).\]
The proof is complete.

We also find that,  the metric $D_G$ is  not joint convex. However,
numerical experiment shows that its square is joint convex, that is,
the following  holds:
\begin{eqnarray}D^2_{G}((\lambda\rho_{1}+(1-\lambda)\rho_{2}),
\sigma)\leq \lambda D^2_{G}(\rho_{1},
\sigma)+(1-\lambda)D^2_{G}(\rho_{2}, \sigma)\nonumber\end{eqnarray}

\section{Conclusion}

In summary, we have introduced a new way to define metric of quantum
states from super-fidelity.  We find that, for qubit case, our
metric $D_G$ coincides with the metric $C(\rho,\sigma)$ introduced
in \cite{Mend}. We proved that the metric $D_{PG}$ is contractive
under quantum operation, while the metric $D_G$ does not behave
monotonically under quantum operation. Also, we rigorously proved
that $D_{PG}$ is joint convex, and numerically proved that the
square of $D_G$ is joint convex. All these show that the metric
$D_G$ is worthwhile studying.

{\bf ACKNOWLEDGMENTS} This work is supported by the New teacher
Foundation of Ministry of Education of P.R.China (Grant No.
20070248087), partially supported by a grant of science and
technology commission of Shanghai Municipality (STCSM, No.
09XD1402500). J.L.C is supported in part by NSF of China (Grant No.
10605013), and Program for New Century Excellent Talents in
University, and the Project-sponsored by SRF for ROCS, SEM.


\begin{thebibliography}{99}

\bibitem{Niebook} M. A. Nielsen, I. L. Chuang, \textit{Quantum Computation and
Quantum Information}, Cambridge Univ. Press, Cambridge, 2000.

\bibitem{Fid1} A. Uhlmann,  Rep. Math. Phys. \textbf{9}, 273(1976);
R. Jozsa,  J. Mod. Opt. \textbf{41}, 2315(1994); M. H\"{u}bner,
Phys. Lett. A \textbf{163}, 239(1992).

\bibitem{97Vedral2275} V. Vedral, M. B. Plenio, M. A. Rippin, and P. L. Knight, Phys. Rev. Lett.\textbf{78}, 2275(1997).

\bibitem{98Vedral1619} V. Vedral and M. B. Plenio, Phys. Rev. A \textbf{57}, 1619(1998).


\bibitem{08Kosut020502} R. L. Kosut, A. Shabani, and D. A. Lidar, Phys. Rev. Lett. \textbf{100}, 020502(2008).


\bibitem{Zana1}
Paolo Giorda and Paolo Zanardi, arxiv: 0903.1262.

\bibitem{Wang1}
Xiaoguang Wang, Zhe Sun, Z. D. Wang, arxiv: 0803.2940;  Xiao-Ming
Lu, Zhe Sun, Xiaoguang Wang, Paolo Zanardi, Phys. Rev. A
\textbf{78}, 032309(2008).

\bibitem{Mis} J. A. Miszczak, Z. Pucha{\l}a, P. Horodecki, A. Uhlmann, and K.
\.Zyczkowski,  Quantum Inf. Comput. \textbf{9}, 0103(2009).

\bibitem{Puch}Z.Pucha{\l}a, J.A.Miszczak, Phys. Rev. A \textbf{79}, 024302(2009).

\bibitem{Mend}
Paulo E. M. F. Mendonca, R. d. J. Napolitano, M. A.Marchiolli,
C.J.Foster, Y.C.Liang, Phys. Rev. A, \textbf{78}, 052330(2008).

\bibitem{Ma1616} Z. H. Ma, F. L. Zhang, D.L.Deng, J. L. Chen,
Phys. Let. A \textbf{373}, 1616(2009).

\bibitem{95Fuchs} C. A. Fuchs, Ph.D. thesis, University of New Mexico, 1995.

\bibitem{Ra02} A.E. Rastegin,  Phys. Rev. A \textbf{66}, 042304(2002).

\bibitem{Nie062310} A. Gilchrist, N. K. Langford, and M. A. Nielsen,
Phys. Rev. A \textbf{71}, 062310(2005).

\bibitem{Ra06} A.E. Rastegin, arxiv: quant-ph/0602112.


\bibitem{Ma064325} Z. H. Ma, F. L. Zhang, J. L. Chen,
Phys. Rev. A \textbf{78}, 064305(2008).


\bibitem{Ma3407} Z. H. Ma, F. L. Zhang, J. L. Chen,
Phys. Let. A \textbf{373}, 3407(2009).

\bibitem{Chen054304} J. L. Chen, L. Fu, A. A. Ungar, X. G.
Zhao,Phys. Rev. A, \textbf{65}, 054304(2002).

\bibitem{Be76} Johan G. F. Belinfante,  J. Math. Phys. \textbf{17}, 285(1976).


\bibitem{Horn91} R. A. Horn, C. R. Johnson,  \textit{Topics in matrix analysis}, Cambridge Univ. Press, Cambridge, 1991.

\bibitem{Bru}
W.Bruzda, V.Cappellini, H.Sommers, K.\.Zyczkowski,  Phys. Lett. A
\textbf{373}, 320(2009).


\bibitem{Pe}
D.P\'{e}rez-Garc\'{i}a, M. Wolf, D.Petz, M.Ruskai,  J. Math. Phys,
\textbf{47}, 083506(2006).

\end{thebibliography}
\end{document}